\documentclass[twocolumn,showpacs,preprintnumbers,amsmath,amssymb,aps,prb]{revtex4}
\usepackage{graphicx}
\begin{document}

\title{
Active Matter Ratchets with an External Drift  
} 
\author{
C. Reichhardt and C. J. Olson Reichhardt 
} 
\affiliation{
Theoretical Division,
Los Alamos National Laboratory, Los Alamos, New Mexico 87545 USA\\ 
} 

\date{\today}
\begin{abstract}
When active matter particles such as swimming bacteria 
are placed in an asymmetric array of funnels, it has been shown that a  
ratchet effect can occur even in the absence of an external drive.  
Here we examine
active ratchets for two dimensional arrays of funnels or L-shapes where 
there is also  
an externally applied dc drive or drift. We show that for certain conditions,
the ratchet effect can be strongly enhanced, and that 
it is possible to have conditions under which 
run-and-tumble particles with one run length 
move in the opposite direction from particles
with a different run length. 
For the arrays of L-shapes, we find that the application
of a drift force can enhance a transverse rectification in the 
direction perpendicular to the drift.
When particle-particle
steric interactions are included, we find that the ratchet effects 
can be either enhanced or suppressed depending 
on barrier geometry, particle run length, and particle density.
\end{abstract}
\pacs{82.70.Dd,83.80.Hj}
\maketitle

\section{Introduction}
When Brownian particles are placed in an asymmetric potential substrate 
in the presence of an external ac drive, it is possible
to realize a so-called ``rocking ratchet'' effect in which the particles
undergo net dc motion \cite{1,2,3}.
Ratchet effects can also be realized
using other forms of external driving such as 
by flashing the substrate on and off to create what is called 
a flashing ratchet \cite{1,2,3}. 
Ratchets have been studied and experimentally realized for 
a variety of systems including colloidal particles on 
asymmetric substrates \cite{4,5}, vortices
in type-II superconductors interacting with nanostructured pinning sites 
\cite{6,7,8}, 
and granular media on vibrated asymmetric substrates \cite{9}.
It is also possible to realize ratchet effects 
on symmetric substrates provided that the external driving has some form
of asymmetry \cite{10,11,12,13,14,15}. More recently, what has been termed 
``active ratchets'' have been
realized in systems where there is no external ac driving or flashing but 
where the particles are self-driven. 
Active matter systems contain self motile particles
\cite{16,17,18,19} and include
biological systems such as swimming bacteria \cite{20,21}, 
moving cells \cite{22,23},
and flocks of birds or fish \cite{V}, as well as  
non-biological systems such as 
artificial swimmers \cite{24,25,26} 
and self motile colloidal particles \cite{27,28,29,30,31}.

In an experiment by Galjada {\it et al.}, when
run-and-tumble swimming {\it E. coli} were placed in a microfabricated array of 
V-shaped funnels, 
the bacteria concentrated on the side of the container towards which the
funnel openings were pointing, indicating the existence of a
ratchet effect \cite{21}. 
When non-swimming bacteria that undergo only weak Brownian motion
were placed in the same funnel array,
the ratchet effect was absent. 
Active ratchet effects have also been observed in funnel geometries
for swimming animals as well as artificial swimmers \cite{35}.     
Subsequent numerical studies showed that this ratchet behavior can be captured
using a model of point particles that undergo run and tumble dynamics along
with a barrier interaction rule stating that when the particles interact with 
a barrier they run along the barrier
rather than reflecting off of it \cite{32}. 
As the run length of the particles is increased, the ratchet effect also 
increases, while in the limit of Brownian
motion the ratchet transport is lost. 
Other studies showed explicitly that the rectification is caused by 
the breaking of detailed balance that occurs when the particles interact
with the barriers, and that the particles must spend a long enough time
running along the barrier for rectification to occur \cite{33,34}.
For other types of barrier interactions such as reflection 
\cite{33,M} or scattering \cite{M}, the
rectification is lost.   
In these simulations it was also   
shown that the particles accumulate in funnel tips and 
along boundaries \cite{33}, a phenomenon that is also
observed in experiments \cite{21}. 
Active ratchets have been studied for other types of swimming 
organisms \cite{36,37,38,39} such as  
crawling cells \cite{23}. In these systems,    
when collective effects are included \cite{38,40,Wan} 
a ratchet reversal can occur where 
for a certain range of parameters
the particles
ratchet along the easy direction of the funnel, 
while for other parameters
the ratchet motion occurs against the easy flow direction \cite{40}.
It was recently proposed that active ratchet effects can arise on 
symmetrical substrates for certain models \cite{41}. 

One of the most promising applications for active ratchets is 
sorting, where different species or particles with 
different run-and-tumble swimming lengths could be sorted due to the different
speed or direction of motion through a ratchet geometry of one type of particle
compared to another \cite{42}.
Variants on this type of ratchet effect have been harnessed to create 
active matter powered gears,
where asymmetric gears immersed in an assembly of active matter 
particles exhibit rotation in a preferred direction 
\cite{43,44}.  There are also proposals to use asymmetric barriers to 
capture active matter particles \cite{H}. 
Another method for sorting rectified Brownian particles is 
to apply a dc drift to the particles that forces them to 
move through a lattice of asymmetric 
obstacles.  In this geometry, particles with different diffusion coefficients 
follow different trajectories
through the array, such that the particle motion perpendicular to the drift
force varies as a function of the diffusion coefficient
\cite{45,46,47}.
This and related methods have been used to continuously sort particles such as 
DNA strands of different lengths using
asymmetric post arrays \cite{47,48,49,49a,50}. 
There have also been several other studies on how to sort 
particles with different diffusive constants in 
periodic arrays when there is an additional dc drift applied 
\cite{50,51,52,53}.

In this work we examine active ratchet systems in which the particles
interact with an array of asymmetric barriers in the presence of an
additional dc drift force.
We consider run-and-tumble particles interacting with two barrier geometries.  
For an array of V shapes or funnels, 
in the absence of a drive the particles exhibit an active ratchet effect
and move in the easy flow direction; however, when a dc drive is applied
against the ratchet effect, we find a ratchet reversal, indicating that
it should be possible to set the dc drive such that particles with different
run lengths move in opposite directions through the funnels.
The velocity-force curves contain nonlinear features 
that vary as the run length changes.
For example, for a fixed dc drive applied against the easy flow
direction, increasing the run length initially increases the flow of
particles in the reversed ratchet direction as the trapping of particles
at the funnel tips is reduced; however, at long run lengths the reversed
motion in the direction of the dc drive is suppressed when the forward
ratchet effect begins to dominate.
We show that the ratchet effect can be controlled by applying a 
dc drive perpendicular to the the ratchet direction.  
Inclusion of steric interactions between particles reduces
the ratchet effect, and the magnitude of the reduction increases as the size
or density of the particles increases.
For an array of even L-shaped barriers, 
application of a dc drive can increase the rectification transverse to 
the applied drive by almost an order of magnitude compared to the
drive-free case.
For this geometry, when steric interactions between the particles 
are included, the transverse ratchet effect is enhanced for
some run lengths and particle densities and suppressed for others.

\section{Simulation and System}   

\begin{figure}
\includegraphics[width=3.5in]{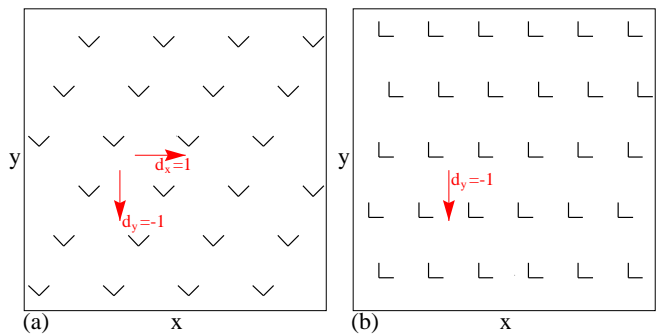}
\caption{ 
(a) Sample geometry for the V shaped barrier array. In the absence of 
an external drive, run-and-tumble particles
ratchet in the positive $y$-direction. 
The arrows indicates the two different directions in which the driving
current can be applied
($d_x=0$, $d_y=-1$ and $d_x=1$, $d_y=0$).
(b) Sample geometry for the even L-shaped barrier array. 
In this case, the dc drive is applied in the negative $y$-direction 
The arrow indicates the direction in which the driving current is
applied
($d_x=0$, $d_y=-1$).
}
\label{fig:1}
\end{figure}

We consider a two-dimensional system of size $L \times L$ 
containing $N$ 
active matter particles
obeying the same rules for run-and-tumble self-propelled motion and 
barrier interactions as previously used to study
ratchets without a drift \cite{32,M}. 
Steric particle-particle interactions are neglected in some sets of
simulations and included in others.
We employ periodic boundary conditions in the $x$ and $y$-directions
for samples containing a periodic array of V-shaped barriers as in
Fig.~1(a) or a periodic array of even L-shaped barriers as in Fig.~1(b).
For the V-shaped barriers, there are $N_B=24$ barriers with side 
length $l_s=5.0$, the V has an angle of $45^{\circ}$, 
and the barrier lattice constant is approximately $a=20$. 
For the even L-shaped barriers, there are $N_B=30$ barriers with 
sides of equal length $l_s=4.9$.
In each case the system size is $L=99$ and there are $N=980$ particles. 
The dynamics of
particle $i$ are obtained by integrating the following overdamped 
equation of motion:
\begin{equation}  
\eta \frac{d {\bf R}_{i}}{dt} = 
 {\bf F}^{m}_{i} + {\bf F}^{i}_{b} + {\bf F}^{s}_{i} +  {\bf F}^{dc}_{i} . 
\end{equation} 
Here the damping constant is $\eta = 1.0$ and   
${\bf F}^{m}_i$ is the motor force. 
The run-and-tumble dynamics is modeled by 
having the particles move with a constant force
$F^{m}$ in a randomly chosen direction
for a fixed run time $\tau_{r}$; after this time, a new running
direction is randomly chosen to represent the tumbling process. 
The tumbling occurs instantaneously.
In the absence of interactions with barriers or other particles, 
a single particle
would move a distance $R_{l} = F^{m}\tau_{r}$ during a single run time. 
The term
${\bf F}_{b}^i$ represents the particle-barrier interaction force. 
The barrier exerts a short-range repulsion on the particle,
modeled by a stiff finite range spring.
As a result, when a particle strikes a barrier it moves along the barrier
at a speed given by the component of its motor force 
that is parallel to the barrier \cite{32,M} until it either reaches the end of
the barrier or undergoes a tumbling event, when it has the opportunity to
move away from the barrier or continue following the barrier at a new speed. 
A particle moving along a barrier can become trapped 
at corners where two barriers meet.
The barrier thickness is equal to the particle radius $R_p$.
The steric interaction between particles, ${\bf F}_i^{s}$, when
included, is modeled
with a repulsive short-range harmonic force given by
${\bf F}^{s}_{i} = \sum_{i\neq j}^N k(R_{eff}^{ij} - |{\bf r}_{ij}|)\Theta(R^{ij}_{eff} - |{\bf r}_{ij}|){\hat {\bf r}}_{ij}$
where the spring constant $k = 200$, 
${\bf r}_{ij} = {\bf R}_{i} - {\bf R}_{j}$, 
${\bf {\hat r}}_{ij} = {\bf r}_{ij}/|{\bf r}_{ij}|$         
and $R^{ij} = r_{i} + r_{j}$, where ${\bf R}_{i(j)}$ is the location of
particle $i(j)$ and 
$r_{i(j)}$ is the radius of particle $i(j)$. 
Here we consider particles of uniform size, $r_i=R_p$.
Unless otherwise noted, we take $R_p=0.35$.
The dc force ${\bf F}^{dc}_{i}=F_{dc}(d_x {\bf \hat x} + d_y{\bf \hat y})$ 
is applied uniformly to all the particles. For $d_x=0$ and $d_y=-1$, 
in the absence of self-driven forces or barriers 
this drive would cause the particles to
drift in the negative $y$ direction. 
We measure
the normalized average particle velocities 
$ \langle V_{x}\rangle = (1/N)\sum^{N}_{i = 1} {\bf v}_{i}\cdot {\hat {\bf x}} $ 
and $\langle V_{y}\rangle = (1/N)\sum^{N}_{i = 1} {\bf v}_{i}\cdot {\hat {\bf y}}$.  

\begin{figure}
\includegraphics[width=3.5in]{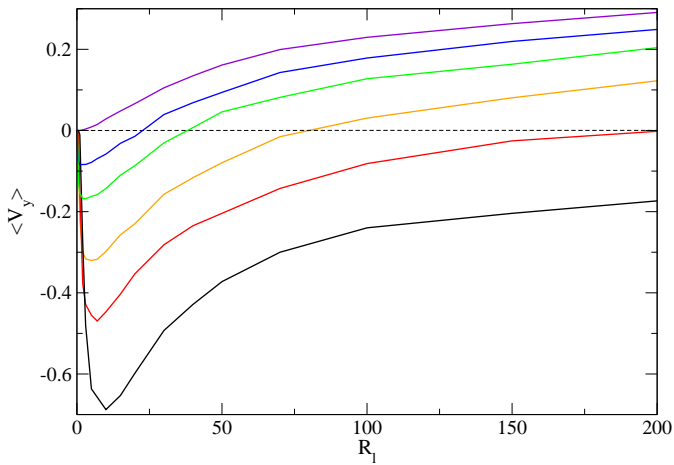}
\caption{ 
$\langle V_{y}\rangle$  vs particle run length 
$R_{l}$ for the V-shaped barrier system in Fig.~1(a) with
$d_x=0$, $d_y=-1$, and
$F_{dc} = 0$, 0.5, 1.0, 2.0, 3.0, and 5.0, from top to bottom. 
At $F_{dc} = 0$ the system exhibits only a positive ratchet effect.  
}
\label{fig:2}
\end{figure}

\section{Funnel-shaped Barriers} 
We first consider non-interacting particles in the array
of V-shaped barriers illustrated in Fig.~1(a).
In the absence of any external
drive, this system shows a rectification effect similar to that observed
for a single row of barriers \cite{32}
where for a finite run length $R_{l}$ there is a net particle current in the 
easy flow directions of the funnels
(positive $y$-direction) which increases with increasing $R_{l}$. 
In Fig.~2 we plot $\langle V_{y}\rangle$ versus run length 
$R_{l}$ for a dc drive applied in the negative $y$ direction
($d_x=0$ and $d_y=-1$), against the
easy flow direction of the funnels.
For each $R_{l}$ we wait a sufficiently long time before measuring 
$\langle V_{y}\rangle$ to avoid any
transient effects. 
Shown in Fig.~2 are the results for
$F_{dc} = 0$, 0.5, 1.0, 2.0, 3.0, and $5.0$. 
For $F_{dc} = 0$, 
$\langle V_{y}\rangle$ is initially zero for $R_{l} = 0$ and monotonically
increases with increasing $R_{l}$, consistent with previous results \cite{32}. 
At finite $F_{d}$
and small $R_{l}$,  
$\langle V_{y}\rangle$ is initially negative and rapidly becomes more negative
as $R_l$ increases until reaching a maximally negative
value between $R_l=1$ and 10,
after which it increases with increasing $R_l$.
For $F_{dc} = 0.5$, 1, and 2, $\langle V_y\rangle$
crosses from negative to
positive $y$-direction flow with increasing $R_{l}$ when the 
positive $y$ direction
ratchet effect becomes large enough to overcome the drift force in the
negative $y$ direction.
This result implies that in a mixed system of particles with short and
long running lengths, there is a range of $F_{dc}$ over which the particles
with short running lengths would move in the negative $y$ direction while
the particles with long running lengths would move in the positive
$y$ direction.
The initial decrease in $\langle V_{y}\rangle$ with increasing 
$R_{l}$ at smaller values
of $R_{l}$ occurs because for small $R_l$ the positive ratchet effect
is weak, and many particles become trapped in the funnel tips due to
the negative $y$ drift force.
For very small $R_l$, most or all of the particles are trapped, 
giving $\langle V_{y}\rangle \approx 0$ 
as shown in Fig.~2. 
As $R_{l}$ increases, some particles can escape from the funnel tip traps but
are then entrained by the drift force to move in the negative $y$ direction,
giving an increasingly negative value of $\langle V_y\rangle$ as more
particles become mobile.
In this regime, increasing the run length
can increase the motion in direction of the drift force;
however, for larger $R_{l}$,
the positive ratchet effect begins to dominate the behavior and
for large enough $R_{l}$ the net flow is in the positive $y$ direction. 

\begin{figure}
\includegraphics[width=3.5in]{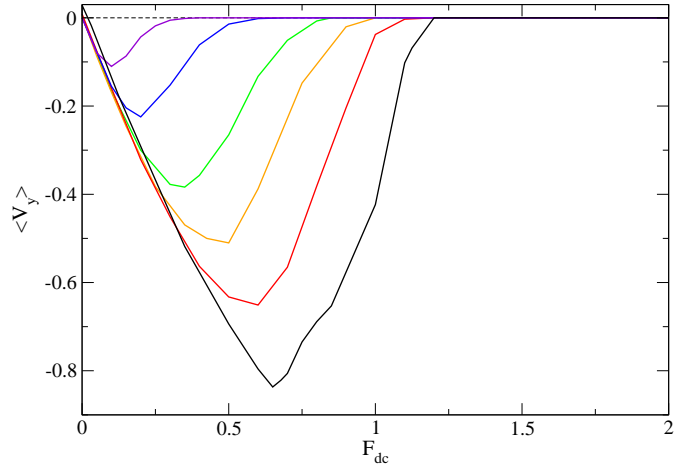}
\caption{
$\langle V_{y}\rangle$ vs $F_{dc}$ for the system in Fig.~2 
with $R_{l} = 0.5$, 1.0, 2.0, 3.0, 5.0, and $10.0$, from top to bottom.
For small $F_{dc}$ the ratchet effect gives a positive
$\langle V_{y}\rangle$.  As $F_{dc}$ increases, $\langle V_y\rangle$ 
becomes negative and then drops to zero as the particles become 
trapped in the funnel tips by the dc drive.   
}
\label{fig:3}
\end{figure}

In Fig.~3 we plot $\langle V_{y}\rangle$ versus $F_{dc}$ for 
$R_{l} = 0.5$, 1.0, 2.0, 3.0, 5.0, and $10$.
At small $F_{dc}$, the ratchet effect 
produces a positive $\langle V_{y}\rangle$. 
For increasing $F_{dc}$,
$\langle V_y\rangle$ crosses zero and becomes negative before reaching
a maximally negative value.
As $F_{dc}$ increases further, particles begin to be trapped in the tips of
the funnels,
and at large enough $F_{dc}$ all the particles
are trapped and $\langle V_{y}\rangle = 0$. 
As $R_{l}$ increases, a larger $F_{dc}$ must be applied for complete 
trapping to occur. 
These results show that the system produces highly nonlinear velocity force 
curves.

\begin{figure}
\includegraphics[width=3.5in]{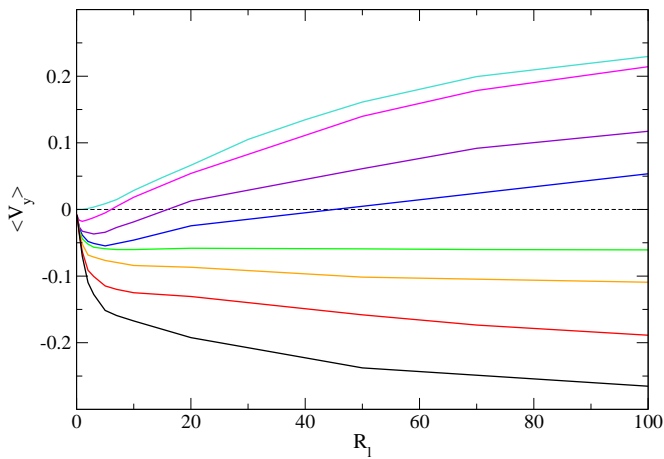}
\caption{
$\langle V_{y}\rangle$ vs $R_{l}$ for the V-shaped barrier system from
Fig.~1(a) with a drive perpendicular to the 
ratchet direction, $d_x=1$ and $d_y=0$, for
$F_{dc} = 0$, 0.5, 1.0, 1.25, 1.5, 2.0, 3.0, and 4.0, from top to bottom. 
For finite $F_{dc}$, $\langle V_y\rangle$ is initially negative for small
values of $R_l$ and becomes increasingly negative
until the ratchet effect becomes strong enough to cause the
particles to move in the positive $y$-direction.  
}
\label{fig:4}
\end{figure}
     
In Fig.~4 we plot $\langle V_{y}\rangle$ vs $R_{l}$ for the same system 
but with $F_{dc}$ applied 
in the positive $x$-direction ($d_x=1$, $d_y=0$), perpendicular 
to the ratchet flow direction.
Here we show $F_{dc} = 0$, 0.5, 1.0, 1.25, 1.5, 2.0, 3.0, and 4.0. 
Due to the barrier shapes, the drift force can alter the motion of the particles
in the ratchet or $y$ direction even though the drift is applied perpendicular
to this direction.
As the particles drift in the positive $x$ direction, they
encounter the outer left surface of a V barrier and follow the barrier
wall downward in the negative $y$ direction before becoming free of the
barrier, encountering another barrier, and again moving in the negative
$y$ direction.
As $F_{dc}$ increases,
the magnitude of $\langle V_{y}\rangle$ rapidly decreases,
and Fig.~4 shows that for $F_{dc}\geq 1.5$, 
$\langle V_y\rangle$ is negative at small $R_l$ but becomes positive for
larger $R_l$ when the ratchet effect becomes strong enough to dominate
the particle motion.
This indicates that with the correct choice of perpendicular dc drive,
particles with different run lengths could be sorted, with one species of 
particles moving in the positive $y$ direction and the other species moving
in the negative $y$ direction.
For $R_{l} = 0$, $\langle V_{y}\rangle$ is zero since the particles
no longer undergo diffusion and end up drifting only in the 
empty horizontal spaces separating adjacent rows of barriers.

\begin{figure}
\includegraphics[width=3.5in]{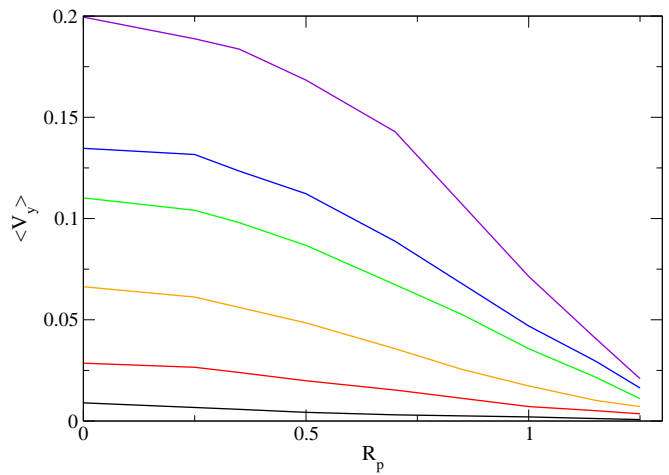}
\caption{
$\langle V_{y}\rangle$ vs $R_{p}$ for the V-shaped barrier system
from Fig.~1(a) with $F_{dc} = 0$ and
steric particle-particle interactions for
$R_{l} = 70$, 40, 30, 20, 10, and $5.0$, from top to bottom.
The ratchet effect decreases monotonically as $R_p$, and the effective
density of the particles, increases.
}
\label{fig:5}
\end{figure}

\begin{figure}
\includegraphics[width=3.5in]{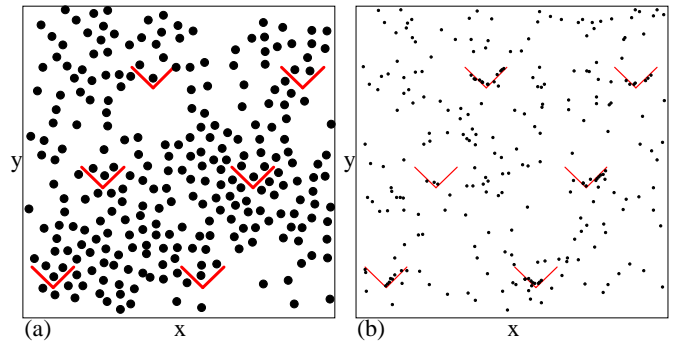}
\caption{
The particle positions (black dots) and barrier locations (red lines)
for a subsection of 
the V-shaped barrier system from Fig.~1(a) with steric particle-particle
interactions.
(a) For $F_{dc} = 0$, $R_{l} = 40$, and $R_{p} = 1.15$, there is a
crowding effect in the traps where at most three particles
can fit in a trap.  Additionally, a clustering effect begins to emerge.
(b) For $F_{dc}=0$, $R_{l} = 40$, and $R_{p} = 0.2$, 
a much larger number of particles can be captured in the funnel tips.    
}
\label{fig:6}
\end{figure}

We next consider the effects of steric repulsion on the ratchet 
effect in the absence
of an external drive. In Fig.~5 
we plot $\langle V_{y}\rangle$ versus the particle radius $R_{p}$ for 
$R_{l} = 70$, 40, 30, 20, 10, and $5$. In each case, inclusion of steric 
interactions causes a drop in the rectification effect due to trapping
and clustering. 
Each funnel can now trap only a limited number of particles, since
particles that would be trapped at the funnel
tip in the noninteracting case instead fill up the funnel, reducing the
trapping effectiveness for particles moving in the negative $y$ direction.
For example,
in Fig.~6(a) we illustrate a subsection of a system with $F_{dc} = 0$,
$R_{l} = 40$, and $R_{p} = 1.15$, where 
at most three particles can fit inside each V shaped barrier
due to the finite size of the particles.
We also observe a clustering of the particles that reduces their overall
mobility.
A similar 
dynamic clustering effect for repulsively interacting active particles 
has been studied as a function of particle density in simulations \cite{L} 
and observed in experiments with active colloids \cite{31}.  
In Fig.~6(b) we show a sample with $R_{p} = 0.2$, where a larger number of 
particles can be trapped in the V barriers. 
This reduces the net downward motion of the particles
since particles traveling in the positive $y$-direction are not trapped by the
barriers.   
As $R_{p}$ increases, fewer particles can be trapped in 
each funnel, and the
net downward motion of the particles increases.
We also find that as the particle radius increases, there is a
decrease in the extent to which the particles are guided
along the sides of the barriers and pushed in the negative $y$ direction,
reducing the rectification.
If two particles are moving along a barrier wall in opposite directions,
the particles can block each others' flow. 
In the system without steric interactions, the particles could instead
pass through each other.
As the particle radius increases, the number of particles that 
can be guided by a given barrier is reduced since fewer particles can fit
on the barrier at the same time. 
An increase in the particle density produces a higher number of 
particle-particle collisions throughout the system, even in the regions
away from the barriers, reducing the effective run length of the particles.
When a drift is applied, an increase in the particle radius
also monotonically decreases the ratchet effect.

\section{L Shaped Barriers}

\begin{figure}
\includegraphics[width=3.5in]{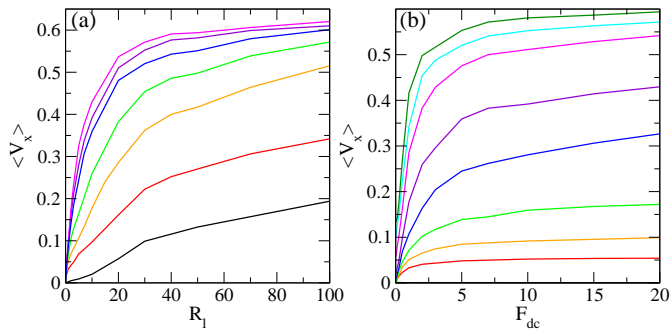}
\caption{$\langle V_x\rangle$, the particle drift in the $x$ direction, 
for the L-shaped barrier system from
Fig.~1(b) with a dc drive applied in the negative $y$-direction
($d_x=0$, $d_y=-1$).
(a) $\langle V_{x}\rangle$ vs $R_{l}$ for $F_{dc} = 0$, 0.5, 1, 2.1, 5, 10, 
and $20$ from bottom to top. 
At $F_{dc} = 0$ there is a ratchet effect in the positive 
$y$ and $x$ directions. As $F_{dc}$ increases,
$\langle V_{x}\rangle$ increases. 
(b) $\langle V_{x}\rangle$ vs $F_{dc}$ for $R_{l} = 0.1$, 0.5, 1, 2, 5, 
10, 20, 30, and $50$, from bottom to top.
Here the effectiveness of the rectification in the $x$-direction 
increases more strongly with increasing $F_{dc}$ for smaller $R_{l}$.     
}
\label{fig:7}
\end{figure}

\begin{figure}
\includegraphics[width=3.5in]{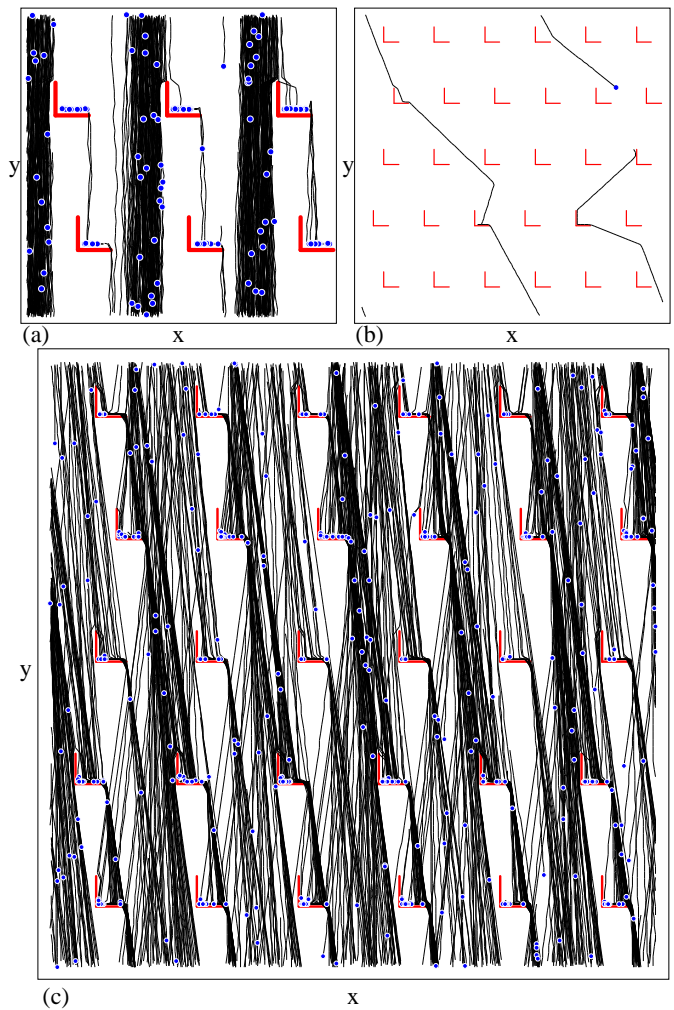}
\caption{
Subsections of the L-shaped barrier system in Fig.~7 
where the dc drive is in the negative $y$-direction, $d_x=0$ and $d_y=-1$. 
Dots: particles; thick lines: barriers; lines: particle trajectories.
(a) For small run length $R_{l} = 0.01$ at $F_{dc} = 5.0$, some particles 
accumulate on the
barriers  while the remaining particles move only in the regions between 
the barriers. (b) The trajectory of a single particle at $R_{l} = 30$ 
and $F_{dc}= 1.0$ shows that over time the particle drifts in the 
positive $x$-direction. 
Additionally,
there are several instances where the particle moves along a barrier and 
is guided to move in the positive $x$-direction.
(c) At $R_{l} = 3.0$ and $F_{dc} = 7.0$,  the trajectories have a net tilt 
in the positive $x$-direction.     
}
\label{fig:8}
\end{figure}

We next consider the even L shaped barriers illustrated in Fig.~1(b).
In this geometry, for $F_{dc} = 0$ and increasing $R_{l}$ the particles 
exhibit a ratchet effect in the positive $y$ and $x$ directions.   
We apply a dc drive in the negative $y$-direction, $d_x=0$ and $d_y=-1$,
and measure the transport in the perpendicular or $x$-direction.
In Fig.~7(a) we plot $\langle V_{x}\rangle$ vs $R_{l}$ for 
systems with $F_{dc} = 0$, 0.5, 1, 2, 5, 10, and $20$.
At $F_{dc} = 0$, $\langle V_{x}\rangle$ increases with increasing $R_{l}$ 
due to the ratchet effect. 
For increasing $F_{dc}$, 
$\langle V_{x}\rangle$ monotonically increases, indicating that a 
transverse ratchet effect occurs.
This is illustrated more clearly in Fig.~7(b)
where we plot $\langle V_{x}\rangle$ versus $F_{dc}$ for
$R_{l} = 0.1$, 0.5, 1.0, 2.0, 5, 10, 20, 30, and $50$.
The effectiveness of the $x$ direction rectification increases the most
rapidly with increasing $F_{dc}$ for the smallest value of $R_l$: 
at $R_{l} = 1$ the ratio of the velocities for 
$F_{dc} = 0$ and $F_{dc} = 20.0$ is nearly $40$,
while at $R_{l} = 5$ it is $4.5$. 
This result indicates that a significant increase in the transverse
ratchet effect can be achieved in active ratchet systems by
applying a drift current. 
For a system of noninteracting particles
with a finite drift force, when $R_{l} = 0$ there is no 
transverse
ratchet effect since the particles either pile up on the barriers or 
flow in the regions between the barriers,
as illustrated in Fig.~8(a) for $R_{l} = 0.01$ and $F_{dc} = 5.0$.
The mechanism by which the dc drift enhances the $x$-direction ratchet effect
is illustrated
in Fig.~8(b) for $R_{l} = 30$ and $F_{dc}= 1.0$, where we highlight the 
trajectory of a single particle.
When the particle encounters the top of a barrier, it can move either
along the outer (left) or inner (right) upper wall of the barrier.
If the particle moves to the inner side of the barrier,
it becomes stuck in the corner of the barrier until it undergoes a
tumbling event that allows it to move away from the barrier in
the positive $x$ direction.  Several
instances of this trap-and-escape motion appear in Fig.~8(b).
If the particle moves to the outer side of the barrier,
it enters the region between barriers and is pushed
in the negative $y$-direction by the dc drive until it encounters another 
barrier, at which point
it can become trapped at the barrier corner before escaping and moving
in the positive $x$-direction. 
This produces a net flux in the
positive $x$-direction
over time, as shown in  Fig.~8(b). 
As $F_{dc}$ is further increased, the particles that move along the 
outside walls
of the barriers into the barrier-free regions
travel more rapidly in the negative $y$ direction and more quickly
encounter additional barriers, 
increasing the effectiveness of the  transverse ratchet effect. 
In Fig.~8(c), the particle trajectories for 
$R_{l} = 3.0$ and $F_{dc} = 7.0$
clearly show a tilt toward the positive $x$-direction.  
We also observe a particle trajectory shadow on the underside
of each barrier.
 
\subsection{Steric Interactions}

\begin{figure}
\includegraphics[width=3.5in]{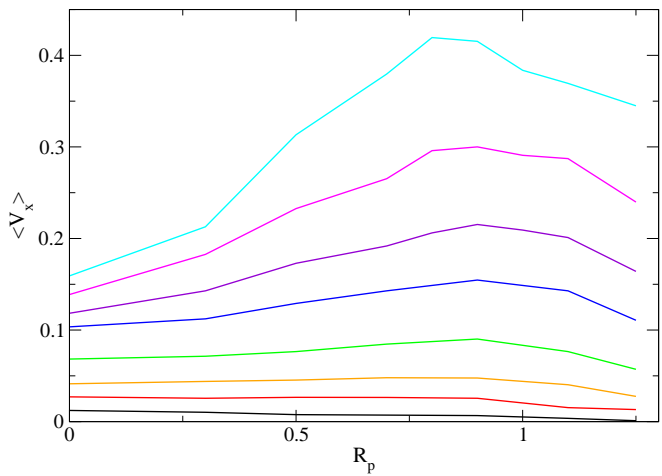}
\caption
{
$\langle V_{x}\rangle$ vs the particle radius $R_{p}$ for
the L-shaped barrier system from Fig.~7 with sterically interacting particles
and $R_l=2.0$ at 
$F_{dc} = 0.1$, 0.25, 0.5, 1, 2, 3, 5, and $10$, from bottom to top. 
For small $F_{dc}$, the steric interactions
reduce the ratchet effect, while for larger $F_{dc}$ there is a 
nonmonotonic response with a peak in $\langle V_{x}\rangle$ at 
$R_{p}\approx 0.8$.
}
\label{fig:9}
\end{figure}

\begin{figure}
\includegraphics[width=3.5in]{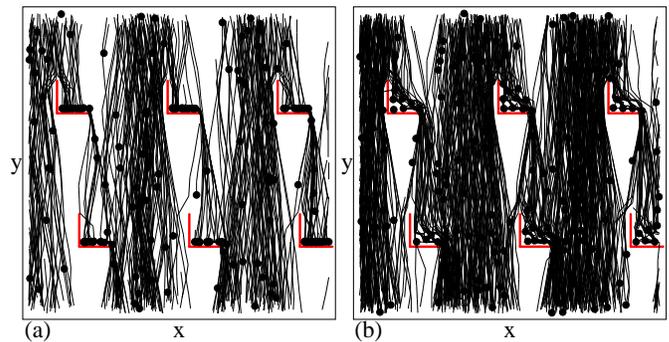}
\caption{ 
Subsections of the L-shaped barrier system in Fig.~9 where the
dc drive is in the negative $y$-direction, $d_x=0$ and $d_y=-1$.
Dots: particles; thick lines: barriers; lines: particle trajectories.
(a) Noninteracting particles with $R_{l} = 2.0$ and $F_{dc} = 5.0$ 
accumulate on the barriers. 
(b) For the interacting particle system with $R_{p} = 0.7$,
$R_l=2.0$, and $F_{dc}=5.0$, fewer particles
are trapped at the barriers.      
}
\label{fig:10}
\end{figure}

We next consider the effects of including steric particle-particle 
interactions for the L-shaped barrier system from Fig.~7.  
In Fig.~9 we plot $\langle V_{x}\rangle$ versus the particle radius 
$R_{p}$ 
for $R_{l} = 2.0$ and $F_{dc} = 0.1$, 0.25, 0.5, 1, 2, 3, 5, and $10$. 
For small or zero $R_{l}$, the addition of steric interactions reduces the 
transverse ratchet effect as also found above for
the V-shaped barriers. 
At higher values of $R_{l}$, however, the steric interactions can 
increase the ratchet effect when $F_{dc} \geq 0.5$, with a 
maximum in $\langle V_{x}\rangle$  occurring at $R_{p} \approx 0.8$. 
In all cases, for $R_{p} > 0.9$ the rectification effect decreases since
fewer particles can fit on the barrier walls to experience guided motion.
The initial increase
in $\langle V_{x}\rangle$ at small $R_p$ for the larger values of $F_{dc}$
occurs due to the filling of the barriers by the particles, as illustrated
for small $R_{l}$ for the non-interacting particles in Fig.~8(a). 
When there are steric interactions,
the number of particles that can be trapped by each barrier is reduced.
In addition,
the trapped particles 
are more likely to move in the positive $x$ direction since
the corner of the barrier becomes blocked by the earliest-arriving particles;
thus, interacting particles that are trapped by a barrier tend to be pushed
to the right in the positive $x$ direction.
Additionally, as particles arrive at the barrier from above,
they fall onto the particles that are already trapped at the barrier and
tend to create a sandpile-like sloped structure with the slope
oriented in the
positive $x$-direction.  
In Fig.~10(a) we plot the particles and their trajectories at $F_{dc} = 5.0$ 
and $R_{l} = 2.0$ for a system without steric interactions, where a pileup of
particles on the barriers occurs.
We show the same system with finite steric interactions and $R_{p} = 0.7$ 
in Fig.~10(b), where we find that fewer particles are trapped
on the barriers due to the
repulsive particle-particle interactions. 
In this case, particles arriving from above the barrier that
interact with
the barrier
tend to be deflected in the positive $x$-direction, 
producing an enhanced rectification.    

\begin{figure}
\includegraphics[width=3.5in]{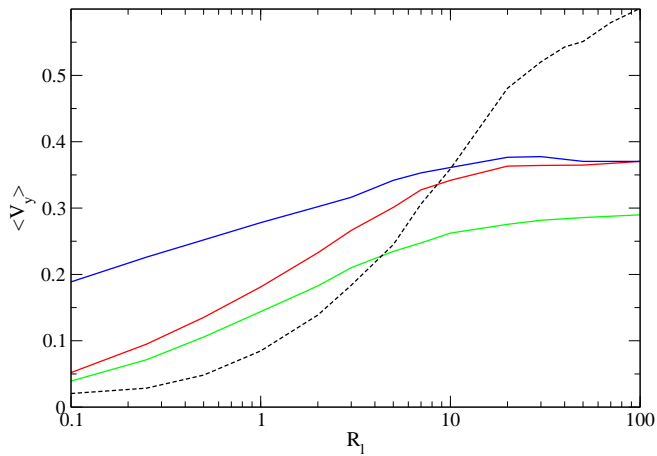}
\caption{
$\langle V_{x}\rangle$ vs $R_{l}$ for the L-shaped barrier system 
with sterically interacting particles from Fig.~9
at $F_{dc}  = 5.0$ for 
$R_{p} = 0.5$, and $R_{p} =  0.9$ with steric interactions
(solid lines from bottom left to top left).
The dashed line is a system with $R_{p} = 0.3$ and no steric interactions. 
Here the particle-particle interactions enhance the rectification
at the lower values of $R_{l}$ and suppresses the rectification
at higher values of $R_{l}$.    
}
\label{fig:11}
\end{figure}

In general, inclusion of steric interactions enhances 
the rectification for shorter 
$R_{l}$, while at longer 
run lengths, the steric interactions decrease the ratchet effect. 
This is more clearly seen in
Fig.~11 where we plot $\langle V_{x}\rangle$ versus $R_{l}$ 
at $F_{dc} = 5.0$ for $R_{p} = 0.3$, 0.5, and 0.9
as well as for a system without steric interactions with $R_p=0.3$.
Here for $R_{l} < 5.0$ the rectification is enhanced by the steric interactions
while for $R_{p} > 5.0$, $\langle V_{x}\rangle$ is higher for the 
noninteracting particles.
At finite $F_{dc}$ and low $R_{l}$, the non-interacting particles 
accumulate in the barriers, reducing
the ratchet effect, while the addition of steric interactions
reduces the number of particles that can interact
with each barrier. 
At larger $R_{l}$ for the noninteracting case,
the particles do not accumulate in the barriers
but there is no limit to the number of particles that can
interact with the barriers, while when steric interactions are
present,
the number of opportunities for particles to interact with the
barriers is limited.

\begin{figure}
\includegraphics[width=3.5in]{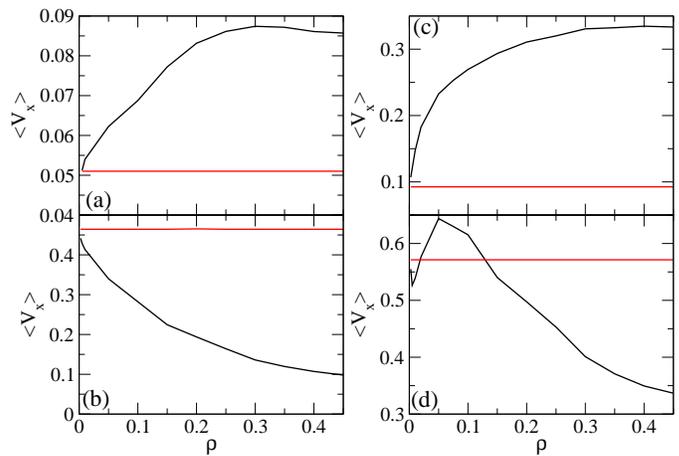}
\caption{
$\langle V_{x}\rangle$ vs $\rho$ for the L-shaped barrier system from 
Fig.~9 with 
sterically interacting particles (curved lines, black) and noninteracting
particles (flat lines, red). 
Here $L$ is held fixed, $\rho = N/L^2$, and $R_{p} = 0.5$. 
In all cases $\langle V_{x}\rangle$ is independent of
$\rho$ for the noninteracting particles.
At small $\rho$ where there are few particle collisions, 
the $\langle V_{x}\rangle$ curves 
for the interacting and noninteracting particles are 
identical or nearly identical.
(a) At $F_{dc} = 1.0$ and $R_{l}  = 1.0$, the ratchet efficiency for the
interacting particle system increases with increasing $\rho$ and is always
higher than in the noninteracting particle system.
(b) At $F_{dc} =1.0$ and $R_{l} = 60$, the ratchet effect 
for the interacting particles is reduced with 
increasing $\rho$ and is always lower than for the noninteracting 
particle system. 
(c) At $R_{l} = 1.0$ and $F_{dc} = 10$, the ratchet effect increases with 
increasing $\rho$ for the interacting particles. 
(d) At $R_{l} = 60$ and $F_{dc} = 10$, the ratchet effect decreases with 
increasing $\rho$ for the interacting particles.    
}
\label{fig:12}
\end{figure}

We can also examine the effects of the steric interactions 
by holding $R_{p}$ and $L$ fixed and varying $N$ to change the particle
density, $\rho=N/L^2$.
For the noninteracting case, $\langle V_{x}\rangle$ is independent  
of $\rho$. 
In Fig.~12(a) we plot $\langle V_{x}\rangle$ versus $\rho$  
for a system with $R_p=0.5$, $F_{dc} = 1.0$ and $R_{l} = 1.0$. 
The result for the noninteracting particles is a flat line.
At low densities where there 
are almost no particle-particle collisions, the value of 
$\langle V_{x}\rangle$ for the interacting and noninteracting systems 
are almost identical.
As $\rho$ increases, $\langle V_{x}\rangle$ increases for the interacting 
particle system until reaching a plateau. 
For values of $\rho$ higher than shown in the figure,
$\langle V_{x}\rangle$ eventually decreases again as the
overall system mobility decreases and the system crystallizes. 
Figure 12(b) shows the
same system with $R_{l} = 60$.  At low densities 
we again find that $\langle V_x\rangle$ for the interacting and noninteracting 
systems are almost the same; however, as $\rho$ increases, 
$\langle V_{x}\rangle$ for the interacting system decreases. 
In Fig.~12(c) we show a system with $R_{l} = 1.0$ and $F_{dc} = 10$, 
where $\langle V_{x}\rangle$ for the interacting system increases with
increasing $\rho$, while Fig.~12(d) shows that for $R_{l} = 60$ and
$F_{dc}=10$,  $\langle V_{x}\rangle$ for the interacting system decreases
with increasing $\rho$. 
These results show that steric interactions 
in combination with a dc drive and short but finite run lengths
increase the transverse ratchet effect, 
while for long run lengths the steric interactions decrease 
the ratchet effect. 
This indicates that it should be possible to use the L-shaped barriers
to sort particles based on both run length and particle radius.

\section{Summary}
We have investigated self-driven particles undergoing run-and-tumble dynamics
in the presence of arrays of V- and L- shaped barriers.  
For the V-shaped barriers in the absence of a drive, we find 
a spontaneous ratchet effect where the particles have a net motion 
in the easy flow direction of the barriers.  
The efficiency of this ratchet effect increases with increasing run length, 
as found in earlier studies of single rows of barriers and in experiments. 
When we apply a dc drift force in the direction opposite to this
ratchet effect, 
we obtain nonlinear velocity-dc force response curves. 
We also observe regimes in which particles
with different run lengths move in opposite directions. 
The introduction of steric particle-particle interactions 
monotonically reduces
the ratchet effect.
For the even L-shaped barriers, which have both arms the same length,
we measure the particle velocity in the direction perpendicular to the 
dc drive and find a transverse ratchet effect that can be substantially
enhanced by the dc drive.
The inclusion of steric interactions can also increase the magnitude
of the transverse ratchet effect.
When the particle radii become too large, this increase is suppressed 
since fewer particles can interact with
each barrier. 
The increase in the transverse ratchet effect
occurs for systems with small but finite run lengths and 
intermediate particle densities
or radii. 
When the run lengths are long, the addition of steric interactions 
generally reduces the ratchet effect. 
Our results show that under a dc drift, active ratchet effects can be
substantially enhanced, 
and provide another approach for controlling the sorting of active matter. 
We also find that steric interactions can in some cases 
produce an increase in the ratchet effectiveness.             

\acknowledgments
This work was carried out under the auspices of the 
NNSA of the 
U.S. DoE
at 
LANL
under Contract No.
DE-AC52-06NA25396.

\end{document}